\documentstyle[aps,prb,multicol,epsf,picinpar]{revtex}

\def\e2 {\epsilon-\epsilon_k}
\def\be {\begin{equation}}
\def\ee {\end{equation}}
\def\bea {\begin{eqnarray}}
\def\eea {\end{eqnarray}}
\def\om {\omega}

\begin{document}
\draft
\title{Interface roughness and planar doping in superlattices:
weak localization effects}
\bigskip
\author{George Kastrinakis}
\address{Institute for Electronic Structure and Laser (IESL), Foundation for
Research and Technology - Hellas (FORTH), \\
Iraklio, Crete 71110, Greece$^*$}

\date{November 29, 2001}
\maketitle
\begin{abstract}

We examine the effects of interface roughness and/or
planar impurity doping in a superlattice, in the frame of
a weak disorder description. We find that these two types
of disorder are equivalent, and that they can be viewed as effective
"bulk" disorder, with anisotropic diffusion coefficients.
Our results offer quantitative insight to transport properties
of multilayers and devices, which contain inadvertently structural
disorder at the interfaces.

\end{abstract}


We consider a superlattice composed of
two materials A and B, each forming layers of thickness
$a$ and $b$ along the $\hat z$ axis. The period of the superlattice 
along the same direction is $d=a+b$.
In a perfect superlattice the electrons feel a periodic potential along the 
$z$ direction,
while they move freely in the $xy$ plane. Hence we 
write the eigenfunctions as products of plane waves in the $xy$ planes times
Bloch functions along $\hat z$ :
\bea
H_o \Psi_{knq} =\epsilon_{knq} \Psi_{knq} \;\;,
\Psi_{knq}({\bf r}) = {e^{i{\bf k \rho}} \Phi_{nq}(z)} \;\;, 
\Phi_{nq}(z) = e^{iqz} u_{nq}(z) \;\;  \label{bloch}
\eea
Here ${\bf k,\rho} \perp \hat z, {\bf q} \parallel \hat z, -{\pi \over d} \leq
q \leq {\pi \over d}, 
\epsilon_{knq} =\epsilon_k + \epsilon_{nq}, \epsilon_k= \frac {k^2}{2m},
\epsilon_{nq}$ being the dispersion relation of the $n$-th band, with a 
bandwidth $w_n$, $n$ running from 1 to $n_b$.
Also, $u_{nq}(z) = u_{nq}(z+ld), l \in N$.
Henceforth we will use $k,K$ and $q,Q$ to denote momenta perpendicular 
and parallel to $\hat z$, respectively.

We consider the effects of scattering either from the roughness of the
interfaces between the two materials forming the superlattice or 
from 'impurities' (dopants) situated in
planes parallel to the interfaces, i.e. vertical to the growth direction $z$.
In this last case, and for charged dopants, we assume that their concentration
is low enough so that the electronic distribution remains unaltered around
the planes. Nevertheless, with a trivial modification in our expressions
for the self energy and the Cooperon \cite{lr} 
our results are also valid for arbitrary kind of low doping.

We only consider spinless disorder, and we model the scattering potential with
\be
V_s({\bf r})= \sum_i V_i({\bf  \rho}) \delta(z-z_i) \;\;,
\ee
where $i$ runs over the interfaces/planes in consideration. $V_i(\bf{\rho})$
describes the roughness/doping profile respectively of the $i$-th 
interface/plane. Henceforth we will use the term 'plane' to denote both.
This type of potential has been used before for the study of 
surface \cite{tesa,fish} and interface \cite{zhang} effects.

Consequently, the Hamiltonian of the superlattice is given by
\be
 H=H_o+V_s \;\;.
\ee
Our perturbative approach consists in treating $V_{s}$ to lowest order. To this
end, we calculate its matrix elements between the states $|knq\rangle$
corresponding
to (\ref{bloch}) above.
For the sake of simplicity we assume that there is only one
scattering plane per period 
(situated in the same relative position c). In fact, this is a very realistic 
possibility, as it often happens that only one of the two interfaces is
rough, e.g. as in GaAs/AlGaAs superlattices.
We will discuss the case of 
two planes per period after deriving the expression for the Cooperon.
We obtain
\be
V_{knq,k'n'q'} =\sum_i V_i({\bf  k' -k})\; \Phi_{nq}^*(c) \Phi_{n'q'}(c) \;\;,
\label{pote}
\ee
where $V({\bf k})$ is the Fourier transform of $V({\bf \rho})$.
There is no correlation between different planes, i.e.
\be
<V_i(k) V_j(k)> = \delta_{ij} <V^2(k)> \;\;.
\ee

 The bare Green's function here is 
\be
G_{on}^{R,A}(k,q,\epsilon)=\frac {1}
{\epsilon-\epsilon_k-\epsilon_{nq} \pm i\delta}, \; \; \;
  \delta \rightarrow 0^+ \;\;.
\ee
The Dyson equation is
\be
G_{nm}=G_{on} \delta_{nm} + \sum_l G_{on} \Sigma_{nl} G_{lm} \;\;,
\ee
where the labels $(k,q,\epsilon)$ are implied. 

To lowest order, 
the self-energy $\Sigma_{nl}(k,q,\epsilon)$ is given by 
\be
\Sigma_{nl}(k,q,\epsilon) = \Phi_{nq}^* \Phi_{lq} \sum_{m,k',q'}
|V(k-k')|^2 G_{om}(k',q',\epsilon) |\Phi_{mq'}|^2  \; \;,
\ee
where we have dropped the argument $c$ from $\Phi$.
In the foregoing we take $V^2=<|V(k)|^2>$.

Setting $\sigma_{nq} = Im \: \Sigma_{nn}(q)$, we assume that
$\epsilon_F - w_n \gg \sigma_{nq}$, which is the condition
$\epsilon_F \tau \gg 1$ in weak localization. We further assume that the
Fermi energy is bigger than the bandwidths $w_n$, in the sense that 
$\epsilon_F-w_n =O(\epsilon_F)$.

	Henceforth we take 
\be
\varepsilon_{nq} = \epsilon_{nq} + \Re \; \Sigma_{nn}(q) \;\;.
\ee

First, we give the Drude part of the static conductivity.
For the in-plane conductivity we have ($\hbar=1$)
\be
\sigma_{xx}= \frac {2e^2}{\pi} \sum_n \int \frac 
{dq \; (\epsilon_F -\varepsilon_{nq})}{\sigma_{nq}} \;\;,
\ee
while for the conductivity along $\hat z$ we have 
\be
\sigma_{zz}= \frac {2\;m\;e^2 }{\pi} \sum_n \int \frac 
{dq \; \epsilon_{nq}'^2}{ \sigma_{nq}} \;\;,
\ee
the prime denoting differentiation with respect to $q$.
Note that this expression, as well as the one for $\delta\sigma_{zz}$
below, behave correctly as the system becomes 2-dimensional. This is the case
in which one material is insulating and forms a thick enough layer.
As the insulating layer becomes thicker the dispersion of the mini-bands
increasingly flattens-out, and hence $\epsilon'_{nq} \rightarrow 0$,
yielding a decreasing conductivity along $\hat z$.

Next, in order to obtain the terms arising from the impurity-induced
electron-hole correlation, we have to calculate the Cooperon first. The
low-field dependence of the conductivity is entirely due to these terms. In
analogy with ref. \cite{tesa} we find for low temperature $T$
\be
C_{q,nmrs}(K,Q,\omega) = V^2 \Phi_{nq}^* \Phi_{mq} \Phi_{r,Q-q}^* 
\Phi_{s,Q-q} ~C(K,Q,\omega) \;\;
\ee
with
\be
C(K,Q,\omega)=\frac {1}{1-V^2 \sum_{k,q,nmrs} G_{nm}^R(k,q,\epsilon_F+
\omega) G_{rs}^A(K-k,Q-q,\epsilon_F) \Phi_{nq} \Phi_{mq}^* \Phi_{r,Q-q} 
\Phi_{s,Q-q}^*} \;\;.
\ee

Assuming that there is reflection symmetry along the $z$ direction,
we evaluate $C(K,Q,\omega)$ for small $K,Q,\omega$ to obtain
\be
C(K,Q,\omega)=\frac {1}{ {\cal D} K^2 +{\cal D}_z Q^2 -i\omega L} \;\;,
\ee
with 
\be
{\cal D} =\frac {t^2}{4m\pi \; S^3} \int dq  
\sum_n \frac {\epsilon_F-\varepsilon_{nq}}
{|\Phi_{nq}|^2}  \;\;, 
\ee
\be
{\cal D}_z = \frac {t}{2\pi S} \int dq 
 \sum_n \frac {1}{\sigma_{nq} } \left[ \frac{|\Phi_{nq}^2|''}{2}
+ |\Phi_{nq}^2| \frac{(\epsilon_{nq}')^2 - (\sigma_{nq}')^2}{8 \sigma_{nq}^2}
- \frac{|\Phi_{nq}^2|' \sigma_{nq}'}{2 \sigma_{nq} } \right]
\;\;, 
\ee
\be
L = \frac {n_b t\pi}{S^2 d} \;\; ,
\ee
where $t=1/2\pi N_2 V^2$,  $S= \frac {1}{2\pi} \int dq \sum_n 
|\Phi_{nq}|^2$. Notice that 
${\cal D}$ and $L$ are equivalent to the quantities $D\tau$ and $\tau$, 
respectively, in weak localization: $D$ being the diffusion coefficient 
and $\tau^{-1}$ the total impurity scattering rate.

If we assume that there is no reflection symmetry along the $z$ direction,
as e.g. in the $(111)$ direction of GaAs \cite{chang} then
$ \epsilon_{n,-q} \neq \epsilon_{nq} $ and $\Phi_{n,-q} \neq
\Phi_{nq}$ in general. As a consequence, a term linear in $Q$ appears in the
denominator of the Cooperon, multiplied by a coefficient $B$, where
\be
B = \frac {i}{2\pi S} \int dq 
 \sum_n \frac {|\Phi_{nq}|^2 \epsilon_{nq}' }{\sigma_{nq} } \;\;.
\ee
However we will not refer further to this case.

Using these expressions, we obtain for the corrections to the conductivity
\be
\delta\sigma_{xx} = - \frac {e^2 V^2}{2\pi^2} 
\sum_{n,m} \int dq \;\frac { |\Phi_{nq}|^2 |\Phi_{mq}|^2}
{(\varepsilon_{nq}-\varepsilon_{mq})^2 + (\sigma_{nq}+\sigma_{mq})^2} 
\; \left \{ \frac{\epsilon_F-\varepsilon_{nq}}{\sigma_{nq}}
+ \frac{\epsilon_F-\varepsilon_{mq}}{\sigma_{mq}}
+3 \; M_{nmq} \right\} \; I \;\; \label{sx}
\ee
\bea
M_{nmq} = \frac { (\varepsilon_{mq}-\varepsilon_{nq}) \; 
(\sigma_{mq}-\sigma_{nq})}
{(\varepsilon_{nq}-\varepsilon_{mq})^2 + 
(\sigma_{nq}-\sigma_{mq})^2} \;, \;m\neq n\;,
\nonumber \\
= \frac{\epsilon_{nq}' \; \sigma_{nq}'}{(\epsilon_{nq}')^2 + (\sigma_{nq}')^2}
\;,\;m=n\;\;, \nonumber
\;\;
\eea
\be
\delta\sigma_{zz} = - \frac {m \; e^2 V^2}{2\pi^2} 
\sum_{n,l} \int dq \; |\Phi_{nq}|^2 |\Phi_{lq}|^2
\frac { \; \epsilon_{nq}' \; \epsilon_{lq}' \; 
(\frac {1}{\sigma_{nq} }+\frac{1}{\sigma_{lq}}) }
{(\varepsilon_{nq}-\varepsilon_{lq})^2 + (\sigma_{nq}+\sigma_{lq})^2}
\; \; I \; \;, \label{sz}
\ee
with
\be
I = \frac{2 L^{1/2} \tau_\phi^{-1/2}(T) }{\pi \sqrt{ {\cal D}^2 \; {\cal D}_z}
} \arctan{\sqrt{\frac{ L \tau_\phi^{-1}(T) }{ {\cal D}_z q_m^2} } }\;\;. 
\ee
$q_m=\pi/d$ and 
$\tau_\phi^{-1}(T)$ is the dephasing rate of the Cooperon \cite{lr}, arising
from electron-phonon, electron-electron interactions etc. $\tau_\phi^{-1}(T)$
enters in the Cooperon in the following manner:
$C(K,Q,\omega)^{-1} \rightarrow C(K,Q,\omega)^{-1} + \tau_\phi^{-1}(T) L$,
where $C$ is given by eq. (14).
$I$ has the limiting forms
\bea
I = \frac {L^{1/2}}{ \sqrt{ {\cal D}^2 \; {\cal D}_z \; \tau_\phi(T) } } \;\;
,\;\; L \tau_\phi^{-1}(T) \gg {\cal D}_z q_m^2 \;\;, \\
I = \frac {2}{ \pi {\cal D} \; {\cal D}_z q_m \tau_\phi(T)  } \;\;
,\;\; L \tau_\phi^{-1}(T) \ll {\cal D}_z q_m^2 \;\;. 
\eea
Weak disorder of bulk type, i.e. spreading over the whole volume, 
in superlattices was investigated e.g. in ref. \cite{sjk}, and
our result for $I$ is the same, in the limit 
$L \tau_\phi^{-1}(T) \gg{\cal D}_z q_m^2$. However, here the prefactors to the 
conductivity corrections as well as the coefficients ${\cal D},{\cal D}_z$
and $L$ are different, due to the fact that we are considering 'planar'
disorder.

Let us note here that the above results for interface scattering were
used in our work \cite{gk} on positive giant magnetoresistance, emanating
from interactions and weak disorder.

a function of the field
\be
I = \frac{\sqrt{e H} }{ 4 \pi^2 \sqrt{ {\cal D} {\cal D}_z }}
\sum_{n=0}^{N_H} \; b_n \; \arctan ( b_n q_o ) \;\;, \;\;
b_n = \left ( \frac{1}{a H}  + n + \frac{1}{2} \right )^{-1/2} \;\;.
\ee
Here $a=4{\cal D}e\tau_{\phi}(T)/L$, $N_H=[q_o^2/\pi^2]$ and
$q_o = \pi/(2 \sqrt{e H {\cal D} })$.
In contrast to this, the 
Drude term has a $(\om_c \tau)^2 \ll 1$ field dependence, with $\om_c$ 
being the cyclotron energy.
Our results are relevant e.g. to the study of Cu/Al multilayers \cite{fad}.

In the case of two planes per period situated in positions $c$ and $c'$,
the potential (\ref{pote}) acquires another identical term with the 
argument $c'$ in $\Phi$. Henceforth we set
\be
\Phi_{1nq}=\Phi_{nq}(c) \;\;,\;\; \Phi_{2nq}=\Phi_{nq}(c') \;\;.
\ee
As a result, the denominator of the Cooperon contains now the constant 
term $1 - A$ with 
\be
A =  \sum_n \int dq \; \frac{
|\Phi_{1nq}|^4+|\Phi_{2nq}|^4+2 |\Phi_{1nq}|^2 |\Phi_{1nq}|^2}
{\sum_l \big \{ |\Phi_{1nq}|^2 |\Phi_{1lq}|^2 + |\Phi_{2nq}|^2 |\Phi_{2lq}|^2
+ 2 \Re \; [ \Phi_{1nq}^* \Phi_{2nq} \Phi_{1lq} \Phi_{2lq}^* ] \big \} } \;\;.
\ee
In this way the Cooperon now is written as
\be
C(K,Q,\omega)=\frac {1}{ 1 - A + {\cal D} K^2 +{\cal D}_z Q^2 -i\omega L
+ \tau_{\phi}^{-1}(T) L } \;\;,
\ee
with 

\be
{\cal D} =\frac {t^2}{4m\pi \; S^3} \int dq  
\sum_n \frac {\{\epsilon_F-\epsilon_{nq} \} [ |\Phi_{1nq}|^4 + |\Phi_{2nq}|^4 ]}
{[ |\Phi_{1nq}|^2 + |\Phi_{2nq}|^2 ]^3}  \;\;, \label{ned}
\ee
\bea
{\cal D}_z = \frac {t}{2\pi S} \int dq 
 \sum_n \frac {1}{\sigma_{nq} } 
\Bigg[ \frac{ |\Phi_{1nq}^2|'' + |\Phi_{2nq}^2|'' }{2}
+ \{ |\Phi_{1nq}^2| + |\Phi_{2nq}^2| \} 
\frac{(\epsilon_{nq}')^2 - (\sigma_{nq}')^2}{8 \sigma_{nq}^2} \\ \nonumber
- \frac{ \{ |\Phi_{1nq}^2|' + |\Phi_{2nq}^2|' \} \;\sigma_{nq}'}
{2 \sigma_{nq} } \Bigg]
\;\;, 
\eea
\be
L = \frac {t}{S^2} \; \int dq \sum_n \frac{|\Phi_{1nq}|^4 + |\Phi_{2nq}|^4}
{[ |\Phi_{1nq}|^2 + |\Phi_{2nq}|^2 ]^2} \;\; . \label{nec}
\ee

Generally, this will result in the incomplete cancellation of the unity 
in the denominator of the Cooperon. 
This remaining term is equivalent to an effective temperature independent
"dephasing rate" or mass term 
\be
\tau_{\phi, eff}^{-1} = \frac{1-A}{L} \;\;.
\ee
If $\tau_{\phi, eff}^{-1} \gg \tau_\phi^{-1}$ the usual weak localization 
signature disappears.
In that case, the diffusive behavior induced by one plane in the unit cell
is washed out by the presence
of the second plane, which in a way 'undoes' the effect of the former.

In summary, we provide an analysis of weak disorder effects from planar
disorder in superlattices, which is equivalent to bulk-type disorder. 
We present and contrast the results from
a single scattering plane per period vs. two scattering planes per period.

\vspace{.8cm}

\vspace{.5cm}
$^*$ E-mail : kast@iesl.forth.gr


\begin{references}

\bibitem{lr}
P.A. Lee and T.V. Ramakrishnan, Rev. Mod. Phys. {\bf 57}, 287 (1985).

\bibitem{tesa}
Z. Tesanovic, M.V. Jaric and S. Maekawa, Phys. Rev. Lett. {\bf 57},
2760 (1986).

\bibitem{fish}
G. Fishman and D. Calecki, Phys. Rev. Lett. {\bf 62}, 1302 (1989).

\bibitem{zhang}
S.Z. Zhang, P.M. Levy and A. Fert, Phys. Rev. B {\bf 45}, 8689 (1992).

\bibitem{chang}
Y.-C. Chang, Phys. Rev. B {\bf 25}, 605 (1982).

\bibitem{sjk}
W. Szott, C. Jedrzejek and W.P. Kirk, Phys. Rev. B {\bf 40}, 1790 (1989).

\bibitem{gk}
G. Kastrinakis, Europhys. Lett. {\bf 42}, 345 (1998).

\bibitem{fad}
A.N. Fadnis, M.L. Trudeau, A. Joly and D.V. Baxter, Phys. Rev. B {\bf 48},
12202 (1993).

\end{references}
\end{document}